\documentclass[12pt]{article}

\usepackage{amsfonts}
 \usepackage{amssymb}
 \usepackage{amsmath}

\def\pmx{\begin{pmatrix}}
\def\emx{\end{pmatrix}}
\def\bsq{\begin{subequations}}
\def\esq{\end{subequations}}

\def\be{\begin{eqnarray}}
\def\ee{\end{eqnarray}}
\def\bee{\begin{eqnarray*}}
\def\eee{\end{eqnarray*}}

\newtheorem{thm}{Theorem}

        \def\pf{\medbreak\noindent{\bf Proof:}\enspace}

             \def\half{{\textstyle \frac{1}{2}}}
 \def\tr{{\rm Tr} \, }
 \def\trp{{\rm Tr} }

\def\wh{\widehat}

\def\bra{\langle}
\def\ket{\rangle}
\def\kb{ \ket \bra }
\def\dg{\dagger}
\def\ot{\otimes}

\def\raw{\rightarrow}

\def\half{{\textstyle \frac{1}{2}}}

\def\dtsig{{\mathbf \cdot {\bold \sigma}}}
\def\bu{{\mathbf{\bold u}}}

\def\bw{{\mathbf{\bold w}}}
\def\bz{{\mathbf{\bold z}}}

\def\bs{{\mathbf{\bold s}}}
\def\bt{{\mathbf{\bold t}}}
\def\b0{{\mathbf{\bold 0}}}

\def\nl{\newline}
\def\nn{\nonumber}

\title{Comments on  multiplicativity of \\
 maximal $p$-norms when $p = 2$}

 \author{Christopher King\thanks{Partially supported
by
  the National Science
        Foundation under Grant DMS-0101205.} \\
 Department of Mathematics\\
Northeastern University,  Boston MA 02115 \\ {\normalsize king@neu.edu}
\and Mary Beth Ruskai\thanks{Partially supported  by
  the National Science
        Foundation under Grant DMS-0314228.}
   \\ Department of Mathematics ,
Tufts University\\
  Medford, Massachusetts 02155 \\ {\normalsize marybeth.ruskai@tufts.edu}}

\date{\today \\ ~~ \\ Dedicated to A.S. Holevo on the occasion of his
60th birthday}

\begin{document}

\maketitle

\begin{abstract}
We consider the maximal
$p$-norm associated with a completely positive map and
the question of its multiplicativity under tensor products.
We give a condition under which this multiplicativity holds when $p = 2$,
and we
describe some maps which satisfy our condition. This class
includes maps for
which multiplicativity is known to fail for large $p$.

Our work raises some questions of independent interest
in matrix theory; these are discussed in two appendices.
\end{abstract}

 \pagebreak

\section{Introduction}

In quantum information theory, noise is modeled by a completely
positive and trace-preserving (CPT) map $\Phi$ acting on the states
of the quantum system.
In general  $\Phi$ takes pure states into mixed states, and
in order to assess the `noisiness' of the map
one is interested in knowing how close the image states may come
to pure states.   Amosov, Holevo and Werner (AHW) \cite{AHW}
observed that this could be measured by the quantity
\be  \label{eq:nu_p}
   \nu_p(\Phi) = \sup\{ ~ \| \Phi(\rho) \|_p ~: \, \rho > 0,
    ~ {\tr} \rho = 1 \}
\ee
where $\| \gamma \|_p =
   \big[ {\tr} (\gamma)^{p} \big]^{1/p}$ and
$1 \leq p \leq \infty$. For any density matrix
$\gamma$, $\| \gamma \|_p \leq 1$ with equality if and only if
$\gamma$ is pure. Hence $\nu_p(\Phi) \leq 1$ with equality if and only
if there is a pure state $\rho$ for which $\Phi(\rho)$ is also pure
(since by convexity the $\sup$ in (\ref{eq:nu_p}) is achieved on a pure state).

The $p$-norm defined above can be extended to arbitrary matrices
as   $\| A \|_p = \big[ {\tr} |A|^p \big]^{1/p}$ with
$|A| = \sqrt{A^{\dg} A}$.
The following useful relationships, which were established in \cite{AH},
can be readily verified.
\be \nu_p(\Phi)
      & = &  \sup_{\gamma > 0, {\tr} \gamma = 1}   \| \Phi(\gamma) \|_p
 = \sup_{A > 0} \frac{\| \Phi(A) \|_p }{ {\tr} A}  \label{eq:nu2} \\
      & = & \sup_{A = A^{\dg}} \frac{\| \Phi(A) \|_p }{{\tr} |A|}
 = \sup_{A = A^{\dg}} \frac{\| \Phi(A) \|_p }{ \| A \|_1}
  \label{eq:nu3} \\  \label{eq:nu4}
   & \leq &
% \sup_{A } \frac{{\tr} (|\Phi(A)|^p)^{1/p} }{ {\tr} |A|} =
 \sup_{A } \frac{\| \Phi(A) \|_p }{ \| A \|_1} .
\ee
(The equivalence of (\ref{eq:nu2}) and (\ref{eq:nu3}) follows from
the convexity of the p-norm and the fact that
$  |\Phi(A_+ - A_-)| = \Phi(A_+) + \Phi(A_-)$ when $A = A_+ - A_-$
%TYPO
is the decomposition of a self-adjoint operator into its positive and
negative parts.)
It follows immediately from (\ref{eq:nu3}) that for any
self-adjoint $A$
\be  \label{eq:nu5}
  \| \Phi(A) \|_p \leq  \nu_p(\Phi) \: {\tr} |A| .
\ee

The representation (\ref{eq:nu4}) suggests viewing $\Phi$ as a map between
spaces of complex matrices with different $p$-norms.
As suggested in \cite{AH} one can generalize this
by defining
\be  \label{eq:norm.qp}
  \| \Phi \|_{q \raw p} =
  \sup_{A } \frac{\| \Phi(A) \|_p }{ \| A \|_q}
\ee
and let $\| \Phi \|_{q \raw p}^R$ denote the same quantity
when the supremum is restricted to the real vector space
of self-adjoint operators.  Then
$\nu_p(\Phi)$ is precisely $\| \Phi \|_{1 \raw p}^R$.
In general, $\| \Phi \|_{q \raw p}^R \leq \| \Phi \|_{q \raw p} $
and one would expect  that the inequality could be strict for
some $\Phi$.  However, the second part
of Theorem~\ref{thm:22} states that equality holds in
all dimensions when $p = q = 2$; and in Appendix~\ref{app:pgeq2.pf},
we show that equality holds for CPT
%TYPO
maps on qubits when $q = 1, p \geq 2$.
This raises the question of whether equality always
holds and, if not, for what types of maps strict
inequality is possible.

%\bigskip

Amosov, Holevo and Werner conjectured \cite{AHW} that $\nu_p$ is
multiplicative on tensor
products, i.e., that
\be   \label{eq:mult}
\nu_p(\Phi \ot \Omega) = \nu_p(\Phi) \, \nu_p(\Omega)
\ee
This has been verified in a number of special cases, although it is now
known to
be false in general.
Amosov and Holevo \cite{AH} proved (\ref{eq:mult}) when
both $\Phi$ and $\Omega$ are products of depolarizing CPT
maps and  $p$ is integer.
King proved (\ref{eq:mult}) for all $p \geq 1$
and arbitrary $\Omega$ under the additional assumption that
$\Phi$ is a unital qubit CPT map \cite{K2}, $\Phi$ is a depolarizing
channel in any dimension \cite{K3}, or $\Phi$ is an entanglement
breaking map  \cite{K4}.
However, Holevo and Werner \cite{HW} also showed
that (\ref{eq:mult}) need {\em not} hold in general
by giving a set of explicit counterexamples for
$p > 4.79$ and $d \geq 3$.

%\medskip
Amosov and Holevo conjectured \cite{AH} that the quantity in
(\ref{eq:norm.qp}) should also be multiplicative for $1 \leq q \leq p$,
i.e., that
\be\label{eq:multp,q}
  \| \Phi \ot \Omega \|_{q \raw p}  =
     \| \Phi \|_{q \raw p}  \, \| \Omega \|_{q\raw p}
\ee
Beckner \cite{B} established an analogous multiplicativity for
commutative systems when $ 1 \leq q \leq p$.
Curiously, Junge \cite{J} proved (\ref{eq:multp,q}) for
completely positive (CP) maps with $p$ and $q$
in the opposite order,
that is for the case $1 \leq p \leq q $.
However,
our main interest is  the case   $q = 1 < p$, and Junge's result does not
seem to shed any direct light on this question.

% \medskip
The conjecture (\ref{eq:mult}) is of greatest interest for $p$ near $1$
since taking the limit as $p \raw 1$ yields the von Neumann entropy of
$\gamma$, another natural measure of purity, and the validity of
(\ref{eq:mult}) for $p$ in an interval of the form $[1,1+\epsilon)$ with
$\epsilon > 0$ would imply additivity of the minimal entropy.
Moreover, it has been shown \cite{Sh} that additivity of minimal entropy
is equivalent to several other important conjectures in quantum information
theory,
including additivity of Holevo capacity and additivity of entanglement of
formation.   Audenaert and Braunstein \cite{AB} have also
observed a connection between multiplicativity for CP maps, and
super-additivity of entanglement of formation.

%\medskip
In view of the Holevo-Werner example,
it is natural to conjecture that (\ref{eq:mult}) holds in
the range $1 \leq p \leq 2$.   This is precisely the  range of
values of $p$
for which the function $f(x) = x^p$ is operator convex, and it
is also the range for which a number of convexity inequalities
hold.   Verifying (\ref{eq:mult}) for the special case of $p = 2$
would suggest its validity for $1 \leq p \leq 2$.   Unfortunately,
even this seemingly simple case is not as straightforward as
one might hope.   In this note, we prove (\ref{eq:mult})
when $p = 2$ for a  special class of CP maps.   Although
this is a rather limited result, it gives some insight
into  the difficulties one encounters in the general case.

Note that the multiplicativity for CPT maps follows if it
holds for all CP maps. We consider
this more general case, as it does not seem more difficult.
In fact, it is not hard to show that multiplicativity (\ref{eq:mult})
holds for all $p \geq 1$ whenever $\Phi$ is an extreme CP map.
This is
because these are precisely the CP maps which can be written in the
form $\Phi(\rho) = A^{\dg} \rho A$, i.e, with one Kraus
operator, and one can then assume without loss of generality
that $A$ is diagonal.   Thus, the extreme CP maps fall into
the ``diagonal'' maps considered in Example 1 below, for
which multiplicativity has been proved.

% \bigskip

% \pagebreak

Although the first part of the following theorem
is included in Junge's
result,  we include an elementary argument here.
\begin{thm}  \label{thm:22}
$ \| \Phi \|_{2 \raw 2}$ is multiplicative, i.e.,
  $\| \Phi \ot \Omega \|_{2 \raw 2} =
     \| \Phi \|_{2 \raw 2} \| \Omega \|_{2 \raw 2}$.
Moreover, $ \| \Phi \|_{2 \raw 2} =  \| \Phi \|_{2 \raw 2}^R$
\end{thm}
\pf  First, recall that the complex $n \times n$ matrices form  a Hilbert
space with respect
to the inner product $\bra A, B \ket = {\tr} A^{\dg} B$, and let
$\wh{\Phi}$ denote the adjoint of the linear operator
$\Phi$ with respect to this inner product.  Since
\be  \label{eq:22pf}
  \big(\| \Phi \|_{2 \raw 2}\big)^2 =
   \sup_A \frac{{\tr} [\Phi(A)]^{\dg}\Phi(A)}{{\tr} A^{\dg} A} =
\sup_{A} \frac{\bra A, (\wh{\Phi} \circ \Phi)(A) \ket}{\bra A, A\ket},
\ee
it follows that
$\| \Phi \|_{2 \raw 2}$ is the usual operator sup norm on
this Hilbert space or, equivalently, the largest singular
value of $\Phi$.  Thus, $\| \Phi \|_{2 \raw 2}$ is the
square root of the largest eigenvalue of $(\wh{\Phi} \circ \Phi)$.
This is the same as the largest eigenvalue of
$(\wh{\Phi} \circ \Phi) \ot I$; therefore,
$\| \Phi \ot I \|_{2 \raw 2} = \| \Phi \|_{2 \raw 2}$.
The main result then follows from the submultiplicativity of
the Hilbert space operator norm under composition since
\bee
     \| \Phi \ot \Omega \|_{2 \raw 2}  & = &
   \| (\Phi \ot I ) \circ (I \ot \Omega )\|_{2 \raw 2} \\
  & \leq & \| (\Phi \ot I )\| ~ \| (I \ot \Omega )\|_{2 \raw 2}
   = \| \Phi \|_{2 \raw 2} \| \Omega \|_{2 \raw 2}.
\eee
\nl Note that since $(\wh{\Phi} \circ \Phi)$ has real eigenvalues
(in fact, they are non-negative), the solutions of the
eigenvector equation $(\wh{\Phi} \circ \Phi)(B) = \mu B$
are self-adjoint (or can be so chosen if $\mu$ is degenerate).
This implies that the supremum in (\ref{eq:22pf}) is
achieved with a self-adjoint $A$, which implies the second statement in the
Theorem.

\section{Main Theorem}

We now find it convenient to introduce some notation.
When $\{ e_j \}$ is an orthonormal basis for
${\bf C}^d$, we will let
$E_{jk} = | e_j \kb e_k |$ denote the matrix with a $1$
in the $j$-th row and $k$-th column and $0$'s elsewhere.
Then the set of operators $\{ E_{jk} \}$ also form
an orthonormal basis for the
$d \times d$ matrices with respect to the Hilbert-Schmidt inner product.
Moreover, if $\Gamma$
is a matrix on ${\bf C}^d \ot {\bf C}^{d^{\prime}} $,
we can write
$\Gamma = \sum_{jk} E_{jk} \ot M_{jk}$ where
$M_{jk} = {\trp_1}  \Gamma (E_{jk}^{\dg} \ot I)$.  This is equivalent
to saying that $\Gamma$ is a block matrix with blocks $M_{jk}$.

If $\Omega$ is a CP map, then
 $(I \ot \Omega)(\Gamma) = \sum_{jk} E_{jk} \ot \Omega(M_{jk}) > 0$
which implies that any $2 \times 2$ submatrix
$\pmx \Omega(M_{jj}) &  \Omega(M_{jk})  \\
   \Omega(M_{kj}) &  \Omega(M_{kk} )  \emx$ is positive
semi-definite. This implies in turn that
\be
\Omega(M_{jk}) = \Omega(M_{jj})^{1/2} \, R_{jk} \,
\Omega(M_{kk} )^{1/2}
\ee
where $R_{jk}$ is a contraction. Hence
\be  \label{eq:blk.schwz}
  {\tr} \Omega( M_{jk}^{\dg}) \Omega( M_{jk}) \leq
  \|\Omega( M_{jj}) \|_2 \, \|\Omega( M_{kk}) \|_2 .
\ee

\medskip

\begin{thm} \label{thm:main}
Let  $\Phi$ and $\Omega$ be CP maps one of which {\em (say $\Phi$)}
satisfies the condition
\be \label{eq:postr}
{\tr} \Phi(E_{ik})^{\dg} \Phi(E_{j \ell})  \geq 0 ~~~~ \forall ~~
i,j,k,\ell .
\ee
Then $\nu_2(\Phi \ot \Omega) = \nu_2(\Phi) \nu_2(\Omega)$.
\end{thm}
\pf Writing an arbitrary density matrix $\Gamma$ as above, one finds
\be
   (\Phi \ot \Omega)(\Gamma) = \sum_{jk} \Phi(E_{jk}) \ot \Omega(M_{jk})
\ee
Thus
\be
\lefteqn{ {\tr}[(\Phi \ot \Omega)(\Gamma)]^{\dg}(\Phi \ot \Omega)(\Gamma)  }
\nn  \\ & = & \sum_{ik} \sum_{j \ell} {\tr} \Phi(E_{ik})^{\dg} \Phi(E_{j
\ell})
  ~ {\tr} \Omega(M_{ik})^{\dg} \Omega(M_{j \ell}) \\  \nn
  & \leq &
 \sum_{ik} \sum_{j \ell} \big|{\tr} \Phi(E_{ik})^{\dg} \Phi(E_{j \ell})
\big|   \| \Omega(M_{ik}) \|_2 \,   \| \Omega(M_{j \ell}) \|_2  \\
   & \leq &
 \sum_{ik} \sum_{j \ell} \big| {\tr} \Phi(E_{ik})^{\dg}
    \Phi(E_{j\ell})\big|
     \sqrt{  \| \Omega(M_{ii}) \|_2 \,  \| \Omega(M_{jj}) \|_2 \,
   \| \Omega(M_{kk}) \|_2  \, \| \Omega(M_{\ell \ell})\|_2 } \nn \\
 & \leq &  [\nu_2(\Omega)]^2  \sum_{ik} \sum_{j \ell} \big| {\tr}
\Phi(E_{ik})^{\dg} \Phi(E_{j \ell}) \big| \sqrt{ {\tr} M_{ii} \,
   {\tr} M_{jj} \, {\tr} M_{kk} \, {\tr} M_{\ell \ell}  }
  \label{eq:absv}
\ee
where we have used (\ref{eq:blk.schwz}) and (\ref{eq:nu5}).  Now, note that
the matrix
\be
 N & = &  \sum_{ik} \sqrt{{\tr} M_{ii} {\tr} M_{kk}} \, E_{ik}
\\ & = &
 \pmx \sqrt{{\tr} M_{11}} \\ \sqrt{{\tr} M_{22}} \\ \vdots \\
   \sqrt{{\tr} M_{dd}} \emx
  \pmx \sqrt{{\tr} M_{11}} & \sqrt{{\tr} M_{22}} & \ldots &
   \sqrt{{\tr} M_{dd}} ~\emx
\ee
is positive semi-definite and
\be
\lefteqn{  \sum_{ik} \sum_{j \ell}  {\tr}
\Phi(E_{ik})^{\dg} \Phi(E_{j \ell})  \sqrt{ {\tr} M_{ii} \,
   {\tr} M_{jj} \, {\tr} M_{kk} \, {\tr} M_{\ell \ell} } }
  \hskip4cm  \nn  \\
 & = &  {\tr} \Phi(N)^{\dg} \Phi(N)   \\
 & \leq & [\nu_2(\Phi)]^2 ({\tr} N )^2 \\
 & = & [\nu_2(\Phi)]^2 \, \Big(\sum_i {\tr} M_{ii} \Big)^2  \nn \\
 & = & [\nu_2(\Phi)]^2 \, ({\tr} \Gamma)^2 .  \label{eq:10}
\ee
When (\ref{eq:postr}) holds
 the absolute value bars are redundant in (\ref{eq:absv}).
One can then
substitute (\ref{eq:10}) in  (\ref{eq:absv}) to yield
\be
 {\tr}[(\Phi \ot \Omega)(\Gamma)]^{\dg}(\Phi \ot \Omega)(\Gamma)
 \leq  [\nu_2(\Omega)]^2 [\nu_2(\Phi)]^2 \, ({\tr} \Gamma)^2 .
\ee
Taking the square root and dividing both sides by
${\tr} \Gamma$, one finds
\be
  \frac{\|(\Phi \ot \Omega)(\Gamma)\|_2} {{\tr} \Gamma} \leq
   \nu_2(\Omega) \nu_2(\Phi) ~~~~~ \forall ~~ \Gamma \geq 0.
\ee
%CHANGE
Taking  the supremum over $\Gamma$ gives the desired result.

% \medskip

Although condition (\ref{eq:postr}) is simple, it is basis
dependent.  In order that $\Phi$ be multiplicative, it
suffices that (\ref{eq:postr}) holds for the matrices
$E_{jk} = | e_j \kb e_k |$ associated with {\em some}
basis for ${\bf C}^d$.   The question of when such a basis
can be found gives rise to some interesting questions in
matrix theory which we remark on in Appendix~\ref{app:pos}.
We remark here only that, although
we do not expect (\ref{eq:postr}) to hold for all CP maps,
we also do not have a counter-example.   Thus, one cannot
exclude the possibility that the hypothesis of
Theorem~\ref{thm:main} is actually satisfied by all CP,
or by all CPT, maps.

\section{Special cases}

It order to show that our results are not vacuous, we now
give some examples of maps which satisfy (\ref{eq:postr}).
\begin{enumerate}

\item  Maps with only diagonal Kraus operators. In this case
$\Phi(E_{jk}) = a_{jk} E_{jk}$ for some positive matrix $A = (a_{jk})$,
and the condition  (\ref{eq:postr}) follows from the orthonormality of the
$\{ E_{jk} \}$. This class of CP maps was studied by Landau and Streater
who named them the diagonal maps. In fact a more complicated analysis using
the Lieb-Thirring inequality can be used to show multiplicativity for all
$p \geq 1$ for these maps \cite{K6}.

\item Maps for which
 $(I \ot \Phi)(M) = \sum_{jk} E_{jk} \ot \Phi(E_{jk})$
has non-negative elements, where $M = \sum_{jk} E_{jk} \ot E_{jk}$
is the maximally entangled state.  (This is the block
matrix with blocks
$\Phi(E_{jk})$; it is sometimes called the Choi matrix or
Jamiolkowski state representative of $\Phi$.)
Although this condition is clearly  sufficient to satisfy
(\ref{eq:postr}), it is not necessary.  For example, let $A$ be a
positive semi-definite matrix with some $a_{jk} < 0$.  Then for that
particular $j,k$ the corresponding map  in Example 1 has
$\Phi(E_{jk}) = a_{jk} E_{jk}$ with one strictly negative element.

\item Multiplicativity at $p=2$ has been proven for all qubit CPT maps
\cite{K1}. However, we can verify the condition  (\ref{eq:postr})
only for a subset of qubit CPT maps. This subset is described using the
parametrization of  qubit maps that was derived in \cite{KR}
and summarized in Appendix~\ref{sect:not}.

In terms of that notation, the condition (\ref{eq:postr})
is satisfied when $t_1 = t_2 = 0$ (since in this case $\Phi(E_{jk})$ is
diagonal
when $j = k$ and skew diagonal when $j \neq k$).  It is
interesting to note that multiplicativity is known to hold for all $p
\geq 1$ under the stronger condition $t_1 = t_2 = t_3 = 0$ \cite{K2}, so
this result suggests that it may hold for $p \geq 1$ for a larger class
of CPT maps.

Another class of qubit maps which satisfy (\ref{eq:postr}) are
those with $\lambda_1 \geq \pm \lambda_2$, $t_2 = 0$, and
$t_1 \geq 0$ (again using the notation in Appendix~\ref{sect:not}).
These maps belong to Example~2
since  $\Phi(E_{jk})$ has non-negative elements for all $j,k$.
Furthermore,
%They also satisfy condition (9) of \cite{K1};
in Theorem~3 of \cite{K1}, King proved that multiplicativity holds
for these channels for all {\em integer} $p \geq 1$, and later
 \cite{K5} extended this to all $p \geq 2$.

\item  The special class of maps satisfying (\ref{eq:form})
and discussed below.  This class includes maps for which
multiplicativity does {\em not} hold for some $p > 2$.

\end{enumerate}

Let $M$ denote a $d \times d$ Hermitian matrix with elements
$m_{jk} = x_{jk} + i y_{jk}$ with $ x_{jk},  y_{jk}$ real.
Let $\Phi: M \mapsto \Phi(M)$ denote a linear map with the
following very special properties:
\be  \label{eq:form}
   [\Phi(M)]_{jk} = \begin{cases}
     \sum_{\ell} d_{j \ell} m_{\ell \ell} & \text{when $ j = k$} , \\
      a_{jk} (x_{jk} + i \epsilon_{jk} y_{jk}) & \text{when $ j \neq k$}
     \end{cases}
\ee
where $d_{j \ell} \geq 0 $, $ a_{jk}$ are the off-diagonal elements of
 a fixed Hermitian matrix and $ \epsilon_{jk} = \epsilon_{kj} = \pm 1$.
The map $\Phi$
is trace-preserving if and only if the matrix $D$ with
elements $d_{j \ell}$ is column stochastic.
Not every
map of the form (\ref{eq:form}) will necessarily be CP.
However, certain special subclasses can be identified.

\begin{itemize}

\item[a)] $a_{jk} = 0 ~~ \forall j \neq k$.  In this case,
$\Phi$ is a QC map consisting of the projection onto the
diagonal part of $M$ followed by the action of a column
stochastic matrix on the classical probability vector
corresponding to the diagonal.

\item[b)] $A > 0$ is a fixed positive semi-definite matrix,
$d_{j \ell} = \delta_{j \ell} a_{jj}$ and $\epsilon_{jk} = + 1$.
This is exactly the diagonal class described above.

\item[c)] $d_{j \ell} = 1 - \delta_{j \ell} $,
$a_{jk} = -1 ~~ \forall ~ j \neq k$ and $\epsilon_{jk} = - 1$.
In this case, $\Phi(M) = ({\tr} M ) I - M^{T}$ and
$\frac{1}{d-1} \Phi(M)$ is the CPT map for which Holevo
and Werner \cite{HW} showed that (\ref{eq:mult}) does {\em not} hold
for large $p$.
\end{itemize}

Since King \cite{K1,K6} showed that multiplicativity holds for all
$p \geq 1$
for maps of type (a) and (b), multiplicativity at $p = 2$
 may not seem  very significant.   However, maps of type (c)
are precisely those used to establish that multiplicativity
does not hold for sufficiently large $p$.
Moreover, the full class includes convex
combinations of maps of type (a) with one of type (b)
or type (c), and King's results do not apply to this class.
Thus this class of maps is neither trivial nor uninteresting.

Although one can verify that CP maps satisfying (\ref{eq:form})
always satisfy the hypothesis of Theorem~\ref{thm:main}, we
state and prove their multiplicativity as a separate result.
Inequality (\ref{eq:blk.schwz})  again plays a key role
in the proof.
\begin{thm}  \label{thm:diag}
Let $\Phi$ be a CP map satisfying (\ref{eq:form}) and let
$\Omega$ be an arbitrary CP map.   Then
$\nu_2(\Phi \ot \Omega) = \nu_2(\Phi) \nu_2(\Omega)$.
\end{thm}
\pf  As before, let $\Gamma = \sum_{jk} E_{jk} \ot M_{jk}$  be
the  matrix with blocks $M_{jk}$.
Then $(\Phi \ot \Omega)(\Gamma) > 0$ has blocks
\be  \label{eq:form1}
   [(\Phi \ot \Omega)(\Gamma)]_{jk} = \begin{cases}
     \sum_{\ell} d_{j \ell} \Omega(M_{\ell \ell}) & \text{when $ j = k$} , \\
      a_{jk} (\Omega( \Re M_{jk}) + i \epsilon_{jk} \Omega( \Im M_{jk})
  & \text{when $ j \neq k$}
     \end{cases}
\ee
where $\Re (M_{jk}) = \half(M_{jk} + M_{jk}^{\dg})$ and
  $\Im( M_{jk}) = \frac{-i }{2}(M_{jk} - M_{jk}^{\dg})$.
A straightforward calculation gives  % \pagebreak
\be
\lefteqn{ {\tr}[(\Phi \ot \Omega)(\Gamma)]^{\dg}(\Phi \ot \Omega)(\Gamma)  }
 \\ & = &  \sum_j \sum_{\ell} \sum_n d_{j \ell} d_{jn}
   {\tr} \Omega(M_{\ell \ell}) \Omega(M_{nn})
  + \sum_{j \neq k} |a_{jk}|^2
    {\tr} \Omega( M_{jk}^{\dg}) \Omega( M_{jk}) \nn \\
  & \leq & \sum_j \sum_{\ell} \sum_n d_{j \ell} d_{jn}
   \|\Omega( M_{\ell \ell}) \|_2 \, \|\Omega( M_{nn}) \|_2
  +  \sum_{j \neq k} |a_{jk}|^2 \|\Omega( M_{jj}) \|_2 \,
    \|\Omega( M_{kk}) \|_2  \nn \\
  & \leq & [\nu_2(\Omega)]^2 \Big( \sum_j \sum_{\ell} \sum_n d_{j \ell}
d_{jn}
    {\tr} M_{\ell \ell} {\tr} M_{nn}  + \sum_{j \neq k} |a_{jk}|^2
   {\tr} M_{jj} {\tr} M_{kk} \Big)   \label{eq:paren}
\ee
where we have used  the Schwarz inequality
for the Hilbert-Schmidt inner product and (\ref{eq:blk.schwz}).
Now,  the term in parentheses in (\ref{eq:paren}) is precisely
${\tr} \Phi(N)^{\dg} \Phi(N) $ where $N$ is the matrix
with elements  $N_{jk} = ({\tr} M_{jj} \, {\tr} M_{kk})^{1/2}$.
The desired result then follows as in
the proof of Theorem~\ref{thm:main} since
\bee
{\tr}[(\Phi \ot \Omega)(\Gamma)]^{\dg}(\Phi \ot \Omega)(\Gamma)
    \leq      [\nu_2(\Omega)]^2 {\tr} \Phi(N)^{\dg} \Phi(N)
     \hskip3cm\\
     \leq    [\nu_2(\Omega)]^2 [\nu_2(\Phi)]^2 \, ({\tr} N )^2
    = [\nu_2(\Omega)]^2 [\nu_2(\Phi)]^2 ({\tr} \Gamma)^2 .
\eee
%which implies the desired result.

\section{Concluding remarks}

If one could replace the operator basis $\{ E_{jk} \}$ by
a more general orthonormal operator basis $\{ G_m \}$
for ${\bf C}^{d \times d}$, one
could always satisfy the analogue of (\ref{eq:postr}).
One need only
choose $\{ G_m \}$ to be the basis which diagonalizes the positive
semi-definite operator  $\wh{\Phi} \Phi$, i.e., for which
\be
  (\wh{\Phi} \circ \Phi)(G_m) = \mu_m^2 G_m .
\ee
where $\mu_m$ are the singular values of $\Phi$.  Then  $
  {\tr} \Phi(G_m)^{\dg} \Phi(G_n) = \mu_n^2 \delta_{mn} \geq 0 $.
Moreover, as noted at the end of the proof of Theorem~1, one can always
choose the basis so that each $G_m = G_m^{\dg}$ is self-adjoint.

Using the orthogonality condition ${\tr} G_m^{\dg} G_n = \delta_{mn}$,
one can show that a density matrix $\Gamma$ on a tensor product space
can be written in the form
\be  \label{eq:gamma}
   \Gamma = \sum_m G_m \ot W_m  \quad \text{with} \quad
  W_m = {\trp_1} \, (G_m^{\dg} \ot I) \Gamma .
\ee
Note  $G_m$   self-adjoint  implies that $W_m$ is also  self-adjoint.

We now try to imitate the proof of Theorem~\ref{thm:main}.
Since $
  (\Phi \ot \Omega)(\Gamma) = \linebreak \sum_m \Phi(G_m) \ot \Omega(W_m)
$,
\be
{\tr}[(\Phi \ot \Omega)(\Gamma)]^{\dg}(\Phi \ot \Omega)(\Gamma) & = &
  \sum_{mn} {\tr} \Phi(G_m)^{\dg} \Phi(G_n) ~ {\tr} \Omega(W_m)^{\dg}
    \Omega(W_n) \nn \\
  & = & \sum_n \mu_n^2  {\tr} \Omega(W_n)^{\dg} \Omega(W_n) \\
  & \leq & [\nu_2(\Omega)]^2 \sum_n \mu_n^2 ~ ({\tr} |W_n| )^2 \\
  & = &  [\nu_2(\Omega)]^2 {\tr} \Phi(N^{\dg}) \Phi(N) \nn \\
  & \leq & [\nu_2(\Omega)]^2 [\nu_2(\Phi)]^2 ({\tr} |N|)^2
\ee
where  $N = \sum_m G_m {\tr} |W_n|$ (and the first inequality implicitly
used the assumption that $G_m$ is self-adjoint so that $W_m$ is).

Unfortunately, we can not conclude that  ${\tr} |N| \leq {\tr} \Gamma$
as needed to complete the proof.
(If we had instead  $N = \sum_m G_m {\tr} W_n$, then we would have
$N = {\trp_2} \, \Gamma > 0$ and ${\tr} N = {\tr} |N| = {\tr} \Gamma$.)
This is a real problem.   Using the
Pauli basis for qubits, for which  $G_k = 2^{-1/2} \sigma_k$,
consider the maximally entangled Bell state
$\Gamma = G_0 \ot G_0 + G_1 \ot G_1 - G_2 \ot G_2 + G_3 \ot G_3$.
Then $\trp_2  \, \Gamma = \half I$, but
$N =  \frac{1}{\sqrt{2}} \pmx 2 & 2 + i \\ 2 - i & 0 \emx$
is not positive semi-definite and $\tr |N| > 1$.

\medskip

Although our results do not prove it, we conjecture that multiplicativity
does hold for
all CP maps at $p=2$. If this conjecture turns out to be false, then it
seems unlikely
that any other value of
$p$ between 1 and 2 would play a special role, and there would probably be
counterexamples to multiplicativity all the way down to $p=1$.
In this case additivity of minimal entropy
would be an isolated result and the attempt to prove it using $p$-norms
would probably
be futile.

% \pagebreak

% \bigskip

\appendix

\section{Comments on positivity condition (\ref{eq:postr}):}
\label{app:pos}

To find conditions under which (\ref{eq:postr}) holds,
note that it is equivalent to the requirement that
\be  \label{eq:Xdef}
  x_{ik,j \ell} = {\tr} E_{ik}^{\dg}  (\wh{\Phi} \circ \Phi) (E_{j \ell})
\geq 0
  ~~~\forall ~ i,j,k,\ell
\ee
which is precisely the condition that the matrix $X$ representing
the positive semi-definite linear operator $\wh{\Phi} \circ \Phi$
in the orthonormal operator basis $\{ E_{jk} \}$ also has
non-negative elements $x_{ik,j \ell}$.

If $|f_j \ket  = U |e_j \ket$ denotes another O.N. basis
for ${\bf C}^d$, then $F_{jk} =  | f_j \kb f_k | = U E_{jk} U^{\dg}$
is an orthonormal operator  basis for ${\bf C}^{d \times d}$,
and the corresponding matrix representative of $\wh{\Phi} \circ \Phi$
is $(U^T \ot U^{\dg}) X (\overline{U} \ot U)$.   In the proof of
Theorem~\ref{thm:main} above we require only that $E_{jk}$
have the form $ |e_j \kb e_k|$
for {\em some} orthonormal basis $\{ e_j \}$ for ${\bf C}^d $
so that  we can use (\ref{eq:blk.schwz}).

Therefore, multiplicativity of $\nu_2(\Phi)$ will hold if
there is a unitary operator $U$ on  ${\bf C}^d $ such that
the matrix $(U^T \ot U^{\dg}) X (\overline{U} \ot U)$
has non-negative elements [where $X$ is the matrix with
elements defined by  (\ref{eq:Xdef})].   Unfortunately,
this is not a very tractable condition.

One way to find an example  of a CP map  satisfying (\ref{eq:postr})
on ${\bf C}^d$ is to find a $d^2 \times d^2$ positive semi-definite
matrix $X$ with non-negative elements.   Regarding the  $d \times d$
blocks of $X$ as $\Phi(E_{jk})$ defines a CP map of
the type considered in Example~2.   However,
as noted before, CP maps satisfying (\ref{eq:postr})
need not have this form.

 A $d^2 \times d^2$ matrix $X$ with
elements $x_{ik,j \ell} \geq 0$ also defines a
positive semi-definite linear map  $\Omega$
which one can write as $\Omega = (\wh{\Phi} \circ \Phi)$.
The map  $\Phi$ then satisfies (\ref{eq:postr}), but
it need not be CP.   We only know that
 $\Phi = \Lambda_U \circ \sqrt{\Omega}$ for
some linear operator $\Lambda_U$ on ${\bf C}^{d \times d}$
which is unitary in the sense
  $\tr \Lambda_U(A)^{\dg} \Lambda_U(B) = \tr A^{\dg} B$
for all $d \times d$ matrices $A,B$.
This implies that $\Phi$
must have the form  $\Phi(A) = U^{\dg} \sqrt{\Omega}(A) U $
for some unitary matrix $U$.\footnote{
The fact that   $\Lambda_U$
must have the form
$\Lambda_U(A) = U^{\dg} A U$
is probably well-known, but was first
brought to the attention of MBR by Nicolas Boulant,
who includes a proof in his paper \cite{Boul}.
If one
writes $\Lambda_U$ in Kraus form, one can then use the fact
that $\wh{\Lambda_U} \Lambda_U = I$ to show that the
Kraus operators can be chosen to be a single unitary.}
  Hence, $\Phi$ is CP if
and only if $\sqrt{\Omega}$ is; however, this does not
seem easy to check.

\section{Qubit maps}  \label{app:cex}

\subsection{Notation:}  \label{sect:not}

It will be useful to summarize some basic facts about the
representation of matrices and CP maps on qubits using
the identity and Pauli matrices as bases.   One can write an
arbitrary matrix as $A = z_0 I + \bz \dtsig$ with $z_0 \in {\bf C},
 \bz = \bw + i \bu$ and $\bw, \bu$ vectors in ${\bf R}^3$.
When $z_0 \neq 0$, any norm satisfies
$ \| z_0 I + \bz \dtsig \| = |z_0| \| I + \frac{1}{z_0} \bz \dtsig \|$.
Therefore, we will present most results only
for $z_0 = 1$; results for $z_0 = 0$  are generally
straightforward.  The most general CP map has the form
\bsq \be  \label{eq:qPhiform}
 \Phi(I + \bz \dtsig) & = & (1 + \bs \cdot \bw + i \, \bs \cdot \bu)  I +
(\bt + T \bw + i T \bu) \dtsig \\
 \Phi(\bz \dtsig) & = & \bs \cdot \bz   I + (T \bz) \dtsig
\ee  \esq
where $\bs, \bt$ are vectors in ${\bf R}^3$ and $T$ is a
real $3 \times 3$ matrix.   $\Phi$ is TP if and only if
$\bs = 0$;  and $\Phi$ is unital if and only if
$\bt = 0$.

As observed in
\cite{KR}, one can use the singular value decomposition to assume
without loss of generality that $T$ is diagonal with
real (but not necessarily positive) elements $\lambda_k$.
This leads to the canonical form
\be   \label{eq:canon}
   \Phi(I + \bw \dtsig) = I + \sum_k (t_k + \lambda_k w_k) \sigma_k
\ee
for CPT maps introduced in \cite{KR}.   Conditions on the
parameters $t_k, \lambda_k$ which guarantee that $\Phi$ is
CPT are given in \cite{RSW}; some special cases were considered
earlier in \cite{AF}.

\subsection{Useful formulas}

We now restrict attention to CPT maps acting on $A =  I + \bz \dtsig$
for which   $\Phi(A) =  I + (\bt + T \bw + i T \bu) \dtsig$.
Then
\be
  A^{\dg} A = (1 + |\bz|^2) I + 2(\bw + \bu \times \bw) \dtsig
\ee
with
\be
  |\bw + \bu \times \bw| = |\bw|^2 + |\bw|^2 \, |\bu|^2
      - (\bu \cdot \bw)^2.
\ee
Therefore,  the eigenvalues of $A^{\dg} A $ are
$1 + |\bz|^2 \pm 2 |\bw + \bu \times \bw|$ or, equivalently,
\be  \label{eq:evalAA}
    1 + |\bz|^2 \pm 2 \sqrt{|\bw|^2 + |\bw|^2 \, |\bu|^2
      - (\bu \cdot \bw)^2}.
\ee
and those of $\Phi(A)^{\dg} \Phi(A)$ are
\be\label{def:lambda}
\phi_{\pm} & =& 1 + |\bt + T\bw|^2 + |T\bu|^2 \nn \\
&&  \pm 2 \sqrt{|\bt + T\bw|^2 (1 + |T\bu|^2) - |(\bt + T\bw) \cdot
(T\bu) |^2}.
\ee
When $(\bt + T\bw) \cdot (T\bu) = 0$, (\ref{def:lambda}) becomes
$\big(  |\bt + T\bw| + \sqrt{  1 + |T\bu|^2 } \big)^2$.

We now wish to evaluate and bound  $ \| A \|_1^2 =(\tr |A|)^2 $
Note that
(\ref{eq:evalAA}) implies that the eigenvalues of
$|A| = \sqrt{A^{\dg} A}$ are
\be
    \sqrt{1 + |\bz|^2 \pm 2 \sqrt{|\bw|^2 + |\bw|^2 \, |\bu|^2
      - (\bu \cdot \bw)^2}},
\ee
and observe that their product can be written as
\bee  \label{eq:sqrtmess}
(1 + |\bz|^2)^2 - 4
    ( |\bw|^2 + |\bw|^2 \, |\bu|^2 - (\bu \cdot \bw)^2 )
    = (1 - |\bw|^2 +  |\bu|^2)^2 + 4 (\bu \cdot \bw)^2   .
\eee
Therefore,
\be
  (\tr |A|)^2 & = &   2\Big( 1 + |\bz|^2 +
   \sqrt{ (1 - |\bw|^2 +  |\bu|^2)^2 + 4 (\bu \cdot \bw)^2 } \, \Big) \\
  & \geq & 2\Big( 1 + |\bw|^2 +  |\bu|^2 +
        \big|1 - |\bw|^2 +  |\bu|^2 \big| \Big) \\
& \geq & \begin{cases}
      4(1 + |\bu|^2)   & \text{if}  ~~ |\bw|^2 \leq 1 + |\bu|^2 \\
      4 |\bw|^2 & \text{if}  ~~ |\bw|^2 > 1 + |\bu|^2 ~.
   \end{cases}   \label{eq:norm1s2} \ee

\subsection{Equality for CPT maps when $p \geq 2$:}  \label{app:pgeq2.pf}

We now show that $ \| \Phi \|_{1 \raw p}^R  = \| \Phi \|_{1 \raw p} $
for CPT maps on qubits when $ p \geq 2$.   For $A = I + \bz \dtsig$
 we prove the somewhat stronger result that
\be  \label{eq:strong}
\frac{\| \Phi[I + (\bw + i \bu) \dtsig] \|_p^2 }
  { \| I + (\bw + i \bu) \dtsig] \|_1^2 } \leq
\frac{\| \Phi[I + \wh{\bw} \cdot {\bf \sigma}) \|_p^2 }{ \| I + \wh{\bw}
\cdot {\bf \sigma} \|_1^2 }
\ee
where $\wh{\bw}$
is the unit vector defined by $\bw = |\bw| \wh{\bw}$.
Our argument will use the following easily verified results.
 When $a \geq 0$ and $m \geq 1$,
  \begin{align} \label{def:f}
f(x) & =    |x + a|^m + |x-a|^m  & \text{is increasing for }~\,  & x > 0~
\,
\\ g(x)  & =   \frac{[f(x)]^{2/m}}{x^2}  &  \text{is decreasing for }~ &
x > 0.
  \label{def:g}  \end{align}
Note that $f(x)$ is symmetric in $x$ and $a$ from which it follows that
the expression on the right in (\ref{def:f}) is also
increasing in $a$.

It follows from (\ref{eq:norm1s2}) and (\ref{def:lambda})  that
\be\label{p-ineq1}
\frac{\| \Phi(A) \|_p^2 }{ \| A \|_1^2 } \leq
 \frac{\Big((\phi_{+})^{p/2} + |\phi_{-}|^{p/2}\Big)^{2/p}}
   {4 \max (|\bw|^2, 1 + |\bu|^2)}
\ee
Since the numerator has the form of $f$ in  (\ref{def:f})
with $m = p/2$, the monotonicity in $a$ implies that
dropping the dot product terms in (\ref{def:lambda})
increases the right side of (\ref{p-ineq1}).  Hence
\be\label{p-ineq2}
\frac{\| \Phi(A) \|_p^2 }{ \| A \|_1^2  } \leq
 \frac{\bigg(\Big(|\bt + T\bw| + \sqrt{1 + |T\bu|^2}\Big)^{p} +
\Big||\bt + T\bw| - \sqrt{1 + |T\bu|^2}\Big|^{p}\bigg)^{2/p}}
   {4 \max (|\bw|^2, 1 + |\bu|^2)}
\ee
Since $|T\bu| \leq |\bu|$, (\ref{def:f}) again implies that
the right side of (\ref{p-ineq2}) increases when $|T\bu|$
is replaced by $|\bu|$ in the numerator.
%TYPO
Also we can only increase the ratio on the right
side  of (\ref{p-ineq2})  by choosing
$\bt \cdot T\bw$ to be positive.  Therefore, we can conclude
from (\ref{def:f}) that this ratio is increasing
in $|\bw|$ for $|\bw|^2 \leq 1 + |\bu|^2$, and
from (\ref{def:g}) that it is decreasing
in $|\bw|$ for $|\bw|^2 \geq 1 + |\bu|^2$. Hence this ratio
 is maximized
when $|\bw|^2 = 1 + |\bu|^2$.
Therefore the ratio in (\ref{p-ineq2}) is less than
\be  \label{eq:bd2}
\frac{\Big(\big(|\bt + T\bw| + |\bw|\big)^{p} +
\big||\bt + T\bw| - |\bw|\big|^{p}\Big)^{2/p}}
   {4 |\bw|^2} ,
\ee
which we want to show is smaller than the RHS of (\ref{eq:strong}).
Since $|\bw|^2 = 1 + |\bu|^2  \geq 1$,
\be
|\bt + T\bw| \leq \big||\bw| \bt + T\bw\big| = |\bw| \, |\bt + T\wh{\bw}|.
\ee
Using (\ref{def:f}) again to replace the $|\bt + T\bw|$ term
in (\ref{eq:bd2}), we find
\be
\frac{\| \Phi(A) \|_p^2 }{ \| A \|_1^2 } & \leq &
\frac{\Big(\big(|\bt + T\wh{\bw}| + 1\big)^{p} +
\big||\bt + T\wh{\bw}| - 1\big|^{p}\Big)^{2/p}}
   {4} \\
& = &
\frac{\| \Phi(I + \wh{\bw} \cdot {\bf \sigma}) \|_p^2 }{ \| I + \wh{\bw}
\cdot {\bf \sigma} \|_1^2 } \\
& \leq & \big( \| \Phi \|_{1 \raw p}^R \big)^2 =
{\nu}_{p}(\Phi)^2 .   \label{eq:finbd1}
\ee

We next consider the case $z_0 = 0$, for which
$A = \bz \dtsig$,  $|A| = |z|I$ and $\Phi(A) = (T \bz) \dtsig$.
Then
  $ \| \Phi(A) \|_p = |T \bz|$ for all $p$ so that
\be   \label{eq:finbd0}
\frac{\| \Phi(A) \|_p  }{ \| A \|_1  } =
\frac{\| \Phi(\bz \dtsig ) \|_p  }{ \| \bz \dtsig \|_1  } =
  \frac{ |T \bz| }{|\bz| } \leq \max_k \lambda_{k} \leq
 \nu_p(\Phi)
\ee
where the last inequality follows from
the fact that
$[\nu_2(\Phi)]^2  \geq \half(1 + |\bt + T\bw| )^2 \geq
   |T\bw|^2$ which can be made equal to $ \max_k \lambda_k^2 $ for
some $\bw$ with $|\bw| = 1$.

To complete the proof, recall that  for $z_0 \neq 0$,
$ \| z_0 I + \bz \dtsig \|_p = |z_0| \, \| I + \frac{1}{z_0} \bz \dtsig
\|_p$ and note that the factor $|z_0|$ will cancel
in any ratio of norms.  Therefore, if we take the supremum over
all complex matrices $A$, we can use (\ref{eq:finbd1}) and
(\ref{eq:finbd0})  to conclude that
$\| \Phi \|_{1 \raw p}  \leq \nu_p(\Phi) =  \| \Phi \|_{1 \raw p}^R$
 when $p \geq 2$.   The reverse inequality
  $\| \Phi \|_{1 \raw p}^R \leq \| \Phi \|_{1 \raw p} $
 always holds; therefore, we must
have equality for $p \geq 2$.

\subsection{Remarks}

Suppose that both $\nu_p(\Phi) = \| \Phi \|_{1 \raw p}^R$ and
$\| \Phi \|_{1 \raw p}$ are multiplicative for some $p, \Phi$.
Suppose also that  $\| \Phi \|_{1 \raw p}^R = \| \Phi \|_{1 \raw p}$
in dimension $d$ (e.g., $d = 2$.)  Then in dimension $d^2$
(e.g., $d = 4$),
\be
  \| \Phi \ot \Phi \|_{1 \raw p}^R = \big(\| \Phi \|_{1 \raw p}^R \big)^2
      = \big(\| \Phi \|_{1 \raw p}  \big)^2 =
    \| \Phi \ot \Phi \|_{1 \raw p}  .
\ee
Thus,  equality also holds in dimension $d^2$ for maps of the
form  $\Phi \ot \Phi$.

The argument in Section~\ref{app:pgeq2.pf} breaks down for
$ 1 \leq p < 2$.   Although one does not expect
(\ref{eq:strong}) to hold, the weaker inequality
$\| \Phi(A) \|_p \leq \|  A \|_1 \nu_p(\Phi)$
might still hold, and this is all that is needed
to show   $\| \Phi \|_{1 \raw p} \leq \| \Phi \|_{1 \raw p}^R $.
However, even for $p = 1$, we have been unable to verify (or find a
counter-example to) this.

\medskip

For the general CP form (\ref{eq:qPhiform}) with
$\bs \neq 0$,~ $\nu_1(\Phi) =  1 + |\bs| > 1$ is achieved with
$\bw = \frac{\bs}{|\bs|}$, and
the eigenvalues of  $\Phi(A)^{\dg} \Phi(A)$
are
\bee   \label{def:lambda.gen}
 (S^2 + |\bt + T\bw|^2 +  |T\bu|^2
   \pm 2 \sqrt{|\bt + T\bw|^2
   [(S^2+ |T\bu|^2] - |(\bt + T\bw)
\cdot (T\bu) |^2}
\eee
with $S^2 = (1  + \bs \cdot \bw)^2 + (\bs \cdot \bu)^2$.
If one tries to use the argument in the previous section,
the RHS of (\ref{p-ineq2}) becomes
\be
\frac{\bigg(\Big(|\bt + T\bw| + \sqrt{S^2 + |T\bu|^2}\Big)^{p} +
\Big||\bt + T\bw| - \sqrt{S^2 + |T\bu|^2}\Big|^{p}\bigg)^{2/p}}
   {4 \max (|\bw|^2, 1 + |\bu|^2)}.
\ee
This does not have the form $|x + a|^m + |x-a|^m  $
because  $a = \sqrt{S^2 + |T\bu|^2} $ and $S^2$
depends on $\bw$.

\bigskip

%\pagebreak

{~~}

\end{document}